\documentclass[reprint,superscriptaddress,twocolumn,amsmath,amssymb,aip,footinbib,jcp]{revtex4-1}

\usepackage{graphicx}
\usepackage{dcolumn}
\usepackage{bm}
\usepackage{amssymb}
\usepackage{amsmath}
\usepackage{nicefrac}
\usepackage{url}

\mathchardef\mhyphen="2D

\begin{document}

\preprint{AIP/123-QED}

% \title[Sample title]{Sample Title:\\with Forced Linebreak\footnote{Error!}}% Force line breaks with \\
% \thanks{Footnote to title of article.}
\title{Redox reactions with empirical potentials: atomistic battery discharge simulations}

% \author{A. Author}
%  \altaffiliation[Also at ]{Physics Department, XYZ University.}%Lines break automatically or can be forced with \\
% \author{B. Author}%
%  \email{Second.Author@institution.edu.}
% \affiliation{ 
% Authors' institution and/or address%\\This line break forced with \textbackslash\textbackslash
% }
% \author{C. Author}
%  \homepage{http://www.Second.institution.edu/~Charlie.Author.}
% \affiliation{%
% Second institution and/or address%\\This line break forced% with \\
% }%
\author{Wolf B. Dapp}
  \email{w.dapp@fz-juelich.de}    
  \affiliation{J\"ulich Supercomputing Centre, Institute for Advanced Simulation, FZ J\"ulich, 52425 J\"ulich, Germany}

\author{Martin H. M\"user} 
  \email{martin.mueser@mx.uni-saarland.de}    
  \affiliation{J\"ulich Supercomputing Centre, Institute for Advanced Simulation, FZ J\"ulich, 52425 J\"ulich, Germany}
  \affiliation{Dept. of Materials Science and Engineering, Universit\"at des Saarlandes, Saarbr\"ucken, Germany}

\date{\today}% It is always \today, today,

\begin{abstract}
Batteries are pivotal components in overcoming some of today's greatest 
technological challenges. Yet to date there is no self-consistent atomistic 
description of a complete battery. We take first steps toward modeling of 
a battery as a whole microscopically. Our focus lies on phenomena occurring 
at the electrode-electrolyte interface which are not easily studied 
with other methods.
We use the redox split-charge equilibration (redoxSQE) method that assigns 
a discrete ionization state to each atom. Along with exchanging partial 
charges across bonds, atoms can swap integer charges. With redoxSQE we 
study the discharge behavior of a nano-battery, and demonstrate that this 
reproduces the generic properties of a macroscopic battery qualitatively. 
Examples are the dependence of the battery's capacity on temperature and 
discharge rate, as well as performance degradation upon recharge.
\end{abstract}

% \pacs{Valid PACS appear here}% PACS, the Physics and Astronomy
%                              % Classification Scheme.
% \keywords{Suggested keywords}%Use showkeys class option if keyword
%                               %display desired

\maketitle

\section{Introduction}\label{sec:intro}

Batteries have been the focus of intense scrutiny in recent years.
This research is driven by specialized requirements for energy storage, e.g.,
capabilities for high current for automotive purposes, 
high capacity batteries to buffer disparities in supply and demand for the power grid, 
and high energy density for batteries used in portable electronic 
devices.\cite{TarasconArmand2001,ArmandTarascon2008,Huggins2009,Linden2010}
Much effort is directed toward studying rechargeable lithium 
batteries,\cite{ScrosatiGarche2010} as that is the preferred 
base material due to the high energy densities achievable, 
its good cyclability properties, a high working voltage,
and its abundance in the Earth's crust. 

Numerical modeling flanks experimental work in the goal to optimize batteries.
Traditionally, macro-homogeneous modeling uses diffusive processes through porous material
to describe batteries on a mesoscale (i.e., smaller than the electrode, 
larger than a molecule), and adds
charge and mass balance equations as well as transfer kinetics across the boundary 
surfaces.\cite{NewmanTobias1962,PollardNewman1981,DoyleEtAl1993,FullerEtAl1994,
DoyleEtAl1996,DarlingNewman1998,BotteEtAl2000,RamadassEtAl2004,SrinivasanNewman2004,
SanthanagopalanEtAl2006,WangEtAl2007,WangSastry2007,ZhangEtAl2007,HarrisEtAl2010}
Such simulations very successfully reproduce the macroscopic behavior of batteries, 
and can be used to optimize parameters such as electrode thickness. 
Mesoscopic porous electrode models form the basis for the Li-ion 
battery module for instance in the commercial COMSOL multi-physics 
simulation package. However, some of the underlying assumptions 
of this approach can be considered ``uncertain at best.''\cite{HarrisEtAl2010}

Mesoscale modeling requires constitutive equations. It is possible to input 
a great many effects based on parameterized experimental data into such descriptions. 
Those include, but are not limited to, 
ionic and electronic conductivities,
specific surface areas,
tortuosities,
porosities,
activity coefficients,
transference numbers,
concentrations,
diffusion coefficients,
current densities,
electrochemical potential,
reaction rate constants,
solid electrolyte interface transport processes,
contact resistances between active material and current collector,
and mechanical properties such as strain tensors,
elastic moduli or fracture strengths.\cite{DoyleEtAl1993,WangSastry2007,ThoratEtAl2009,HarrisEtAl2010,AwarkeEtAl2012}
Unfortunately, this parameterizability which makes the approach suitable 
to describe real batteries also limits 
its general predictive power, especially on the nanoscale. 

In recent molecular dynamics (MD) works, on the other hand, much effort has been devoted to dealing with
certain aspects of battery design in detail. Examples are plentiful and include 
improving the intercalation of lithium into graphite, 
or calculating the transport properties of lithium through various electrode materials.
\cite{AbouHamadEtAl2010,LiuCao2010,FisherEtAl2008,EllisEtAl2010,Islam2010,HautierEtAl2011,TripathiEtAl2011,LeePark2012,OuyangEtAl2004,ArrouvelEtAl2009}
In contrast to mesoscale models, MD simulations tackle micro- and nano-physical 
aspects of the problem (typically only considering half-cells), 
but cannot make macroscopic predictions on, 
for instance, how the voltage during discharge changes under the influence 
of various control parameters such as temperature, discharge current, 
or changes of cell geometry. 

Standard MD methods --- based on either conventional \textit{ab initio} density functional 
theory (DFT) or traditional \textit{charge equilibration} (QE)~\cite{MortierEtAl1985} ---
cannot be used to model a battery as a whole, because those seek to 
equalize the chemical potential. However, the difference in the chemical potential 
between two electrodes is precisely what drives charge transport in a battery. 
Furthermore, both approaches 
are ill-suited to model history-dependent effects. The reason is that they carry out 
a unique energy minimization based on instantaneous nuclear positions. The electron 
transfer process during a redox reaction brings about a quasi-discontinuous change of 
the electronic state,  modifying all molecular orbitals,\cite{MayKuehn2011} while 
the atomic configuration remains virtually unaltered. 

Time-dependent DFT (TDDFT)~\cite{RungeGross1984,GrossKohn1990,BauernschmittAhlrichs1996}
can elucidate history dependence to some degree.
However, this is expensive computationally, and, more severely,
conceptual difficulties remain.
They pertain to (a) setting up a meaningful initial
state with the correct voltage between anode and cathode, and (b) 
reproducing correct level hopping due to an overestimation of the 
long-range polarizability in current DFT schemes.  The application 
of TDDFT, and similarly \textit{ab initio} MD, to electrochemical processes
with regards to batteries has been limited.\cite{BlumbergerEtAl2005,MaoEtAl2012}

Recently, it was proposed that the limitations of MD calculations can be mitigated by
introducing the oxidation state as a time-dependent variable, which needs
to be subjected to dynamics.\cite{Mueser2012,VerstraelenEtAl2012,DappMueser2013a}
In addition to exchanging partial charges as in the standard \textit{split charge 
equilibration} (SQE) method,\cite{NistorEtAl2006}
atoms can then change their ionization state (i.e., participate in a redox reaction) 
by swapping integer charges across a bond (integer charge transfer -- ICT). 
Hereafter, the method is referred to as the redox split-charge equilibration 
(redoxSQE) method.

In a previous work,\cite{DappMueser2013a} we applied redoxSQE to case studies of contact 
electrification between two clusters of ideal metals and ideal dielectrics, 
respectively. If two initially neutral clusters with differing electron 
affinities are brought into contact, they will exchange charge. 
After separation, some portion of the transferred charge does not flow back, generating a 
remnant electric field that was not present before the contact. Neither conventional QE 
methods nor (non-time-dependent) DFT can capture this history-dependence.  
RedoxSQE in contrast, successfully produces charge hysteresis effects during approach and 
retraction, despite identical atomic positions. 

In this paper we apply the redoxSQE method to a more complex problem.
We want to bridge the gap between the mesoscale and highly accurate (DFT, 
accurate force field) approaches, and model a nano-battery. 
If properly parameterized, redoxSQE can be used to model the microphysics 
at both electrode-electrolyte interfaces, including their structural evolution 
and changing morphology, as well as battery performance degradation.
At this stage, the simulations are meant to serve as proof-of-concept, rather
than emulate any real system or produce new quantitative insights.
However, even at its present qualitative level, our model reproduces generic 
features of battery discharge. 
We believe that redoxSQE can be parameterized to describe real materials quantitatively, 
because unmodified SQE combined with REBO (reactive empirical bond-order) force fields has yielded good 
agreement with experimental and DFT results (heats of formation of isolated molecules,
radial distribution functions for water and ethanol, and energies of oxygenated diamond surfaces) 
for systems in which each element had a well-defined oxidation state.\cite{MikulskiEtAl2009, KnippenbergEtAl2012}

This paper is structured as follows. We outline the method in Sec. 
\ref{sec:method} below. We also introduce the additional parameters 
and procedures not covered in Ref.~\cite{DappMueser2013a}, 
which describes the method in greater detail. Sec.
\ref{sec:Setup} covers the setup of the specific simulations in this work. 
In Sec.~\ref{sec:results} 
of this paper, we present the results attained by varying both internal and 
external parameters, and compare the outcome to 
generic properties of macroscopic batteries. We close with a discussion 
and summary of our findings in Sec.~\ref{sec:discussion}.

\section{Method}\label{sec:method}

This section briefly outlines the numerical methods (redoxSQE) used in 
this study, and the parameters involved. For a more detailed description 
we refer the reader to Ref.~\cite{DappMueser2013a}. For 
comparisons with DFT-based results see the work by Verstraelen 
\textit{et al.},\cite{VerstraelenEtAl2012,VerstraelenEtAl2013} whose SQE+Q$^{0}$ is similar
in spirit to redoxSQE, but applies charge constraints to fragments of 
molecules rather than to alter the oxidation state of individual atoms.

We implement molecular dynamics with a long-range potential due to fractional charges (``split charges''), 
as discussed in Ref.~\cite{NistorEtAl2006} We add in the modification proposed in Ref.:~\cite{Mueser2012}
\begin{align}
  V_{\mathrm{total}} &= V_{\mathrm{short}} + V_{\mathrm{long}}, \\
  V_{\mathrm{long}} &= V_{\mathrm{C}} 
    + \sum_{i} \left( \frac{\kappa _{i}}{2} Q_{i}^{2}
    + \chi _{i}Q_{i}\right) 
    + \sum_{i, j, j<i} \frac{\kappa _{ij}^{(\mathrm{b})}}{2} q_{ij}^{2}, \label{eq:SQE_potential} \\
  Q_{i} &= n_{i}e + \sum _{j} q_{ij}. \label{eq:Q_from_SQs}
\end{align}
In this expression, $V_{\mathrm{short}}$ represents the short-ranged potential (see below), 
$V_{\mathrm{C}}$ is the standard Coulomb potential, while the $\chi _{i}$ are the 
electronegativities, and the term involving $\kappa _{i}$ is due to the atomic hardness 
(as in the standard QE \cite{RappeGoddard1991}).
The total atomic charges, $Q_{i}$, are the sum of an integer charge
(in $n_{i}$ increments of the elementary charge $e$) on an individual atom, 
as well as partial charges $q_{ij}$ that are shared between 
any two bonded atoms (see Secs. \ref{subsec:BondHard} and \ref{subsec:SQE}). 
The fractional charges are antisymmetric in their indices, i.e., 
$q_{ij} \equiv -q_{ji}$. Single subscripts refer to quantities on 
individual particles (e.g., total charges or atomic properties), 
while double subscripts refer to quantities shared between two 
particles, such as a split charge, or a bond property. 

The last term of Eq.~(\ref{eq:SQE_potential}) describes the effect of the bond hardness 
(as also used in the \textit{atom-atom charge transfer} (AACT) framework~\cite{ChelliEtAl1999}), 
and is discussed in detail below.
% or the MM3 force-field~\cite{DosenMicovicEtAl1983a,DosenMicovicEtAl1983b}). 
Parameterizing the potential in terms of \textit{both} atomic and bond properties alleviates 
most issues that methods only containing one or the other suffered from (see Ref.~\cite{DappMueser2013a} 
and references therein, for a summary of SQE's advantages over 
other \textit{electronegativity equalization methods}, which we do not repeat here). 
% Note that the 
% bond hardness depends sensitively on the distance between the bonded atoms. The parameter 
% $\kappa _{ij}^{(\mathrm{b})}$ is discussed in detail 
% % in Section~\ref{subsec:BondHard} 
% below. 

The equations of motion are solved with a conventional velocity-Verlet algorithm, in a dedicated MD code. 
We use a Langevin thermostat~\cite{SchneiderStoll1978} combined with stochastic damping, 
with a damping constant of $\gamma \Delta t = 1/600$ after the initial equilibration.
The Coulomb interaction is effected in a na\"{\i}ve $\mathcal{O}(N^{2})$ direct-sum approach
(see Sec.~\ref{subsec:limitations} below).

For simplicity, we use the ``6-12'' Lennard-Jones (LJ) potential 
without cutoff for the short-range interactions.
The electrolyte is a Kob-Andersen-like mixture,\cite{KobAndersen1995} in order to prevent crystallization. 
We use six atom types, two for the electrolyte (positively and negatively charged), one for each 
electrode in its neutral state, and one for
each ionized electrode species. Electrode atoms have twice the mass of solvent (electrolyte) atoms. 
Our choice of parameters is summarized in Table~\ref{table:LJ_parameters}. The unit system is 
explained in Sec.~\ref{subsec:units}.
% We set up a two-dimensional system, confined by walls, and with the electrodes on either end, and 
% a separator in the center
% (described in Sec.~\ref{subsec:Separator}). 

\subsection{Bond hardness}\label{subsec:BondHard}

The bond hardness $\kappa _{ij}^{(\mathrm{b})}$ between particles $i$ and $j$, 
which are a distance $r_{ij}$ apart,
is parameterized by the following piecewise function:
\begin{align}
  \kappa _{ij}^{(\mathrm{b})} = \left\{
	\begin{array}{ll}
		\kappa^{(\rm p)}_{ij}	& r_{ij} \le r_{\mathrm{s}},\\
		\kappa^{(\rm p)}_{ij}+\kappa^{(0)}_{ij} \frac {r_{\mathrm{l}}^{2}\left(r_{ij}-r_{\mathrm{s}}\right)^{2}}
                                         {r_{\mathrm{s}}^{2}\left(r_{\mathrm{l}}-r_{ij}\right)^{2}}
			& r_{\mathrm{s}} < r_{ij} < r_{\mathrm{l}},\\
		\infty	& r_{\mathrm{l}} \le r_{ij},\\
	\end{array}
	\right.    	
   \label{eq:BondHard}
\end{align}
where $r_{\mathrm{s}}$ ($\ne 0$) and $r_{\mathrm{l}}$ are short and long cutoff radii, respectively.
The symbols $\kappa^{(\mathrm{p})}_{ij}$ and $\kappa^{(0)}_{ij}$ denote bond parameters, constant 
for each bond type. In this paper the plateau value $\kappa^{(\mathrm{p})}_{ij} = 0$, because we 
are primarily concerned with metallic contacts. 
The functional form of the bond hardness is similar to that of 
Mathieu.\cite{Mathieu2007} However, while we work with two variable 
critical radii, that work parameterized only the inner threshold 
with a variable scaling factor $\lambda_{ij}$. 
Note that Mathieu shortly simplifies 
$\lambda_{ij}\equiv \lambda$ as universal, i.e., not only as bond-independent but also atom-independent. 
The other parameter in Ref.~\cite{Mathieu2007} is a multiplicative factor $C_{ij}$. 
This effectively incorporates our $\kappa^{(0)}_{ij} r_{\mathrm{l}}^{2} / r_{\mathrm{s}}^{2}$. 
Mathieu uses the van-der-Waals radius of a given atom instead 
of a variable outer threshold $r_{\mathrm{l}}$.

Our parameterization of the bond hardness smoothly approaches 
$\kappa^{(\mathrm{p})}_{ij}$ at the lower threshold, while it 
diverges at the upper threshold, where the bond breaks. At both thresholds, the force 
brought about by the distance-dependence of the bond hardness has a cusp, which may lead to very 
small drifts in the total energy. This is discussed in detail in the previous work.\cite{DappMueser2013a}

\subsection{Split charge equilibration (SQE)}\label{subsec:SQE}

% This section outlines our implementation of bond-based partial charges into the simulation. 
% The earlier paper~\cite{DappMueser2013a} summarizes a number of references comparing the (unmodified) 
% SQE method with other popular charge equilibration schemes which showcase SQE's advantages over 
% other \textit{electronegativity equalization methods} (EEMs), which we will not repeat here.
% In the equilibration step (prior to calculating the Coulomb force, and the MD step), 
% we update all split charges on ``active'' bonds for a fixed atomic configuration. 

Prior to calculating the Coulomb force and the MD step, in the 
so-called equilibration step, we update all split charges on 
``active'' bonds for a fixed atomic configuration.
A bond is classified as ``active'' if its bond hardness is zero or finite (but not 
infinite), i.e. if the bond length $r_{ij} < r_{\mathrm{l}}$. 
Inactive bonds do not carry partial charges. 
We minimize the 
potential energy with respect to the split charge distribution by solving the 
homogeneous linear system of equations
\begin{align}
  \partial V / \partial q_{ij} = 0
\label{eq:SQE}
\end{align}
with a steepest-descent solver.\cite{numericalRecipes} The potential $V$ is that of 
Eq.~(\ref{eq:SQE_potential}). Typically, the minimization requires only a handful of iterations. 
However, following an integer charge transfer (see Sec.~\ref{subsec:ICT}), up to 
several thousand iterations can be necessary to find the split charge distribution minimizing the energy. 

Next, we update the total charge on each atom according to Eq.~(\ref{eq:Q_from_SQs}), 
based on the atomic integer charges carried, and the bond charges, and proceed
with the normal MD calculation. 

For the battery simulations shown below, we only allow split-charge 
(and integer-charge) exchange between electrode atoms. Electrolyte 
atoms are modeled as fixed-charge particles.

\subsection{Integer charge transfer (ICT) in dielectric bonds}\label{subsec:ICT}

The novel feature of redoxSQE is that it allows for integer charge transfer (ICT) besides the 
exchange of partial charges across dielectric bonds.\cite{DappMueser2013a}
This section briefly describes the implementation, while  
Sec.~\ref{subsec:ICTM} explains how we treat charge transfer across metallic bonds.

At each time step, we select all ``dielectric'' bonds with $r_{\mathrm{s}} <  r_{ij} < r_{\mathrm{l}}$, 
i.e., any pair of atoms that is sufficiently close together to share a split charge, but not close enough 
to have a vanishing bond hardness (``metallic'' bonds). 
We also exclude electrolyte atoms from participating in ICTs because they are assumed to be unreactive 
with the electrode for maximum battery efficiency.\cite{ArmandTarascon2008,Linden2010}
% Furthermore, we check that a change in oxidation state is possible in the first place, or likely 
% (e.g., no double ionization of hydrogen is possible, and Cu has a high energy threshold to be ionized triply). 
Furthermore, our na\"{i}ve implementation sets hard limits on the 
oxidation state for each atom type --- in the simulations presented below
we do not allow double ionization. 
More sophisticated rules are conceivable, but left for future work.
 
% After the equilibration of the split charges (see Sec.~\ref{subsec:SQE}), 
% we compute the system's total potential energy.
For each eligible bond, we draw a random number between zero and one, uniformly distributed. 
If it is smaller than a certain threshold (somewhat arbitrarily chosen $\equiv 1/\kappa _{ij}^{(\mathrm{b})}$), 
an ICT is attempted. This is to approximate the electron transfer rate. In more
realistic simulations, this rate must be determined from 
quantum-chemical calculations.
For a trial ICT, we increment or decrement the integer charge 
(i.e., the oxidation state) of each participating atom by one elementary charge, 
with the algebraic sign the same as the sign of the split charge between the two atoms. 
Then we re-equilibrate the partial charges, and calculate the system's total potential energy.
If the charge transfer has lowered the energy (i.e., the system now evolves on a 
Landau-Zener level with strictly lower energy), the move is accepted, otherwise 
it is rejected and the original state restored. A modification in future code 
implementations will be be to accept ICTs according to some Metropolis-type condition 
instead.\cite{numericalRecipes} Then, the energy can also increase with a certain probability 
during an ICT, fulfilling the principle of detailed balance, and
producing the correct equilibrium distributions. 
In our current model we expect that the Metropolis algorithm
mainly changes the dynamics near the transition state, i.e., the
reorganization of the solvent might take longer due to back jumps.
However, the final state will not be altered because fluctuations
of solvated ions to become neutral are extremely rare events.

Besides changing the oxidation state, an ICT also changes the atom type. 
This is necessary because an ion may have different atomic properties 
(such as radius and interaction parameters) as well as bond characteristics 
from its neutral counterpart. The type change necessitates also taking
into account the short-range interaction energy for the ICT.
An electrode atom only changes its type if it gains the ``correct'' charge. 
The opposite charge is absorbed by, and distributed across, metallic bonds, e.g., 
among connected remaining electrode atoms. For the anode, this means that an atom is stripped of a 
negative integer charge (i.e., one or more electrons), and becomes a cation 
(i.e., it changes from atom type 1 into type 3), while the 
negative charge remains on the anode. It serves to compensate for the positive charge 
accumulated by sending negative charge through the external resistor. While the net 
positive charge is localized on the cation, the remnant negative charge is 
distributed across the entire anode as split charges instantaneously, even though 
\textit{formally} there is still \textit{one particular} anode atom that carries 
the charge via its oxidation state, for book keeping. Conversely at the cathode, 
adsorbed cations receive negative integer charges transferred from the anode and are neutralized 
(they change from atom type 4 to type 2), as their surplus net positive charge is absorbed.

In addition to the procedure described above, we draw 
another random number and only proceed with the trial ICT 
if this exceeds some threshold (for instance $0.9$, to attempt an ICT
only every tenth MD step), in order to alleviate a 
bias introduced by the order in which we query bonds. This results in
two trial ICTs per atomic oscillation period, on average. In future
implementations we will randomize the order of bonds for which we
attempt an ICT, and fully eliminate the bias.

All in all, the maximum number of attempted ICTs per MD time step is $\sim kgNZ$, where 
$N$ is the number of redox-active atoms, $Z$ is their average coordination number, 
and $k$ is the fraction of ICTs that passed picking the second random number (e.g. 
$10\%$). Lastly, $c^{2} < g < c$, where $c\le 1$ is the fraction of dielectrically 
bonded atoms. The factor is $g \approx c^{2}$ for atoms dissolved in redox-inert 
solvent (e.g., in the electrolyte), and $g \lesssim c$ in clusters of redox-active 
material. 

\subsection{Diffusion of oxidation state: integer charge transfer in metals (ICTM)}\label{subsec:ICTM}

For a ``metallic'' bond with $\kappa _{ij}^{(\mathrm{b})} \equiv 0$, the backflow 
of partial charge exactly compensates the transfer of integer
charge. Such a move would always be accepted because the energy is unchanged, 
but would still cause an expensive yet unnecessary re-equilibration of split charges. 
In addition to the integer charge transfer across a ``dielectric'' bond with 
finite bond hardness, we therefore implement a second mode of ICT that applies 
to metallic bonds. We call such an operation ICTM. 

We implement ICTMs such that we draw a random number for each metallic bond
between two atoms otherwise eligible of an ICT. If 
this number exceeds a threshold, an integer charge is swapped, and immediately 
compensated by an equal split charge transfer in the opposite direction. Together, 
those transfers are energetically neutral moves in a metal. No SQE needs to be 
performed, and no type change occurs, so no further computations are needed. 
ICTMs thereby allow for ``oxidation state diffusion.'' 
If the integer charge get transferred onto the ``front atoms'' in the anode
(the atom connected to the wire, see Sec. \ref{subsec:ext_resistor} below), we do not 
allow it to move away anymore.\footnote{The condition of maximum oxidation 
state remains to be enforced, but is modified for the front atom to read that 
its \textit{effective} oxidation state cannot be outside $\pm 1$, i.e., the 
total charge after adding the ``external'' split charge (connected via resistor 
to the other electrode).} As a consequence, all free negative integer charges
(which could be interpreted as electrons) eventually migrate to the front atom, and are 
sent across the external resistor, in accordance with the real physical process. 
Similarly, negative integer charges emanate from the cathode's front atom and diffuse 
toward cations adsorbed to the electrode surface. 

ICTM happens as next-neighbor 
hopping. It would be more meaningful
if a metal cluster as a whole had an excess of integer charge (positive or negative), 
rather than individual atoms in a metal cluster being assigned an oxidation state. 
This would also reproduce realistic physics more faithfully by allowing an immediate 
transfer of integer charges between any two atoms connected to a metallic cluster.
However, for bookkeeping and domain decomposition reasons, we stick to the current 
procedure, eliminating the overhead of a cluster analysis.

We emphasize that during an ICTM, only negative integer charges can diffuse through 
the electrodes. It is \textit{not} possible for two initially neutral metal atoms 
to assume the configuration +1/-1, in contrast to the ICT in dielectrics, because 
one of the two atoms changes its type in such a case.

Also note that the random number is necessary to reduce (albeit not fully avoid) 
spurious directed motion. We perform the scan for the ICTM deterministically 
(e.g., atom $i$ is always queried before atom $i+1$), and therefore 
would introduce a preferred transfer order if every move was accepted. 

\section{Simulation setup}\label{sec:Setup}

\begin{figure}[hbtp]
  \includegraphics[width=1.0\columnwidth,angle=0]{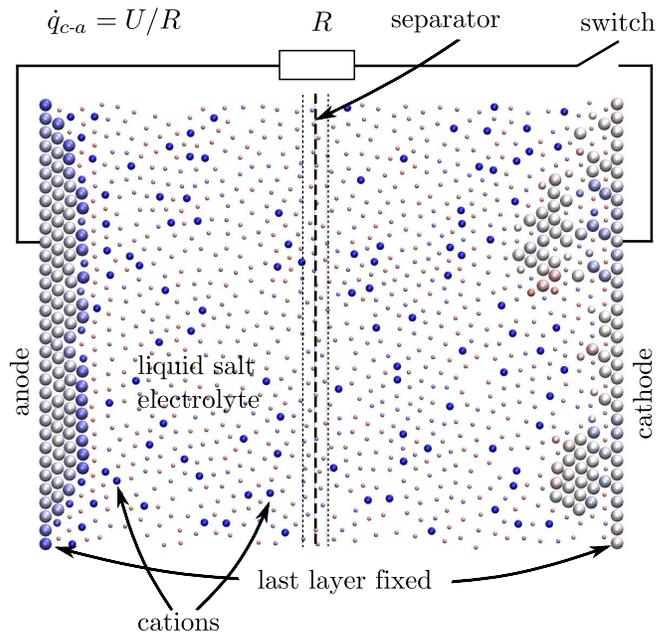}
  \caption{ Color online. Illustration of a setup with 1194 atoms. Charge is encoded in the coloring, 
            blue meaning positive charge and red negative charge. For visual distinction, electrolyte 
            particles are chosen smallest, independent of their LJ radii, and their charge coloring is halved.
            The medium-sized particles are cations, while the largest particles are metallic atoms.
            The separator keeps non-electrolyte atoms from moving between the two half-cells.
            A resistive external load $R$ (following Ohm's law) completes the circuit. 
  }
  \label{fig:snapshot}
\end{figure}

The configuration of a simulation with 1194 atoms at an intermediate time is qualitatively 
illustrated in Fig.~\ref{fig:snapshot}. 
(Our default setup is somewhat smaller, and described in Sec.~\ref{subsec:Setup}.)
The total charge on a given atom is encoded in 
its color, blue being a positive charge and red a negative charge. For visual distinction, 
atoms are displayed at sizes that do not reflect their LJ radii. Both metallic and ionic species
are visualized as much bigger than electrolyte atoms are. The latter are +1/-1 fixed-charge 
particles, but their charge 
coloring is halved, again for better contrast. The medium-sized particles are cations, 
while the largest particles are metallic atoms. Those may or may not carry a negative 
oxidation state, as negative integer charges can hop freely across metallic bonds in an ICTM (see 
Sec.~\ref{subsec:ICTM}). 

The anode in this simulation initially held 5 layers of atoms; all atoms of the rightmost layer have
undergone redox reactions, each donated a negative integer charge to the anode, and subsequently dissolved 
as a cation (roughly 45 atoms in total in this picture). The other layers are still in the 
middle of this process, and are only partially dissolved. The anode is positively charged, 
partly as a result of polarization 
charges as consequence of the layer of anionic electrolyte particles that wet its surface, 
and partly because insufficient atoms have dissolved as cations to carry away its excess charge.
As expected for a metal, the charges reside primarily on the bulk surface~\cite{DappMueser2013a}. 

The cathode started out with only the single fixed layer at the beginning of the simulation, 
but in this snapshot, about 75 pre-dissolved cations in the right half-cell have already 
adsorbed, donated their charge, and become part of the metallic cathode. This process is not homogeneous, the 
accretion occurs in clusters and produces an uneven surface with fractal-like features.
If redoxSQE is properly parameterized for real materials, it can be useful in the quest to 
understand the detailed morphology and structure of the surface layer, as well as phenomena 
such as dendrite formation,\cite{ValovEtAl2011} because it resolves the 
electrode-electrolyte interface. It may also be useful for intercalation studies of lithium into graphite. 

Besides some minor polarization charges, the 
cathode is neutral overall because adsorbed 
cations have contributed sufficient charges to compensate for the negative charge that has
traveled along the connection through the external resistor $R$ 
(see Sec.~\ref{subsec:ext_resistor}). 

The external load is implemented
as a dedicated partial charge between an atom in the fixed layer of either electrode. 
This split charge between cathode and anode $q_{c\mhyphen a}$ is not updated in the SQE 
step (see Sec.~\ref{subsec:SQE}) but receives its value according to Ohm's law (if 
the switch is closed):
\begin{align}
  \dot{q}_{c\mhyphen a} = U/R,
\end{align}
where $R$ is the constant external resistance, and $U$ is the instantaneous driving voltage
between the two connected front atoms. The separator only allows electrolyte 
particles to pass through, all others experience a repulsive force (see 
Sec.~\ref{subsec:Separator}) if they move in between the two thin lines.

\subsection{Initial setup and equilibration}\label{subsec:Setup}

We briefly describe the initial configuration of our default 
setup, and the start of a simulation. 

\begin{itemize}
\item We work in a two-dimensional configuration, to reduce the computational 
      costs associated with large particle numbers. We do not apply periodic 
      boundary conditions: our finite system is confined by fixed walls on all sides.      
\item Our default system contains 358 atoms overall. We have also carried out 
      simulations with roughly double and quadruple that number. 
\item Of the total number of atoms, 118 are electrode atoms. Of those,
      half are attached to the anode and are arranged 
      in an hexagonal lattice with three (up to five, in larger simulations) layers.
      Only 20 atoms are connected to the cathode in one layer, another 39
      ionized cathode atoms are dissolved in the electrolyte. This represents a crude approximation 
      to solubility equilibrium. In a historic Voltaic cell, the dissolved particles would be Cu$^{2+}$ ions.
\item The electrolyte is distributed spatially randomly in the two half cells, under the constraint that 
      each side is initially electrically neutral.
\item We choose our MD time steps such that we sample a typical thermal oscillation of an atom with $\approx 20$ steps.
\item Initially, we allow the electrolyte to equilibrate in each half-cell, barring all atoms from passing the separator. 
      During this stage, the electrode particles are held stationary, and no ICTs/ICTMs are carried out.
\item After $5,000$ MD time steps, the electrolyte has assumed a liquid glass-like state. We insert 
      the separator instead of the impassable wall between the two half-cells, and allow both ICTs and ICTMs.
\item We equilibrate for another $10,000$ MD time steps with all electrolyte atoms free to move, 
      but damped at $\gamma \Delta t = 1/60$, a factor $10$ more strongly than our default value.
\item Finally, the electrode atoms are released, and the damping is set to its normal, low value. 
      Only the last row of each electrode remains fixed in place. 
      This limits the amount of charge transferable to roughly
      66\% of the electrode.
\item We measure during the following $\approx 10^{7}$ MD steps.
\end{itemize}

\subsection{Separator}\label{subsec:Separator}

In order to isolate the two half-cells of our battery, a simple model of 
a \textit{separator} is inserted. In Galvanic cells, this component (called 
``salt bridge'' in that context) is a membrane often made of filter paper, or 
consists of a U-shaped glass tube filled 
with (possibly gelified) inert electrolyte.\cite{Linden2010}
It allows ionic species to pass through, and thus to complete the circuit with the external resistor, but 
prevents intermixing of dissolved electrode ions, which is often undesired. 

In a present-day, commercial batteries, the two half cells are kept apart 
by a solid separator which is permeated by the electrolyte. In the ideal case, 
no electrons are allowed to pass through, but the ion conductivity is large. 
The separator also needs to be mechanically resistant to abuse,
chemically stable in a concentrated alkaline environment (for alkaline 
batteries), as well as not participating in redox reactions that occur in 
the cell.\cite{Linden2010}

We implement a mathematical separator such that dissolved electrode atoms and ions feel 
a repulsive force when they approach the barrier, but electrolyte particles 
are unaffected (in our model they carry charge, and can complete the circuit). 
If electrode particles were allowed to mix, they could exchange split charges 
and form salts, or even adsorb to the opposing electrode and thus create a 
short circuit.

Notice that the energy barrier posed by the separator is not infinitely high. 
As a thermally-activated process that occurs with a probability of 
$\exp\left(-E_{\mathrm{sep}}/k_{\mathrm{B}}T\right)$, individual ions are 
expected to still pass the barrier. We choose the separator's repulsive energy 
$\equiv4$ in dimensionless units ($\approx 0.6~\mathrm{eV}$). This means that an ion 
sitting at the separator has a probability to pass through of 
$\approx 4.5\times 10^{-5}$ at our default temperature.

Our separator is a crude idealization of its real-world counterpart, even though 
their properties are similar. In a more realistic and properly parameterized 
simulation, the separator is a crucial component in its own right, and needs 
to be implemented carefully.

\subsection{External circuit}\label{subsec:ext_resistor}

One atom of the fixed layer on each electrode is chosen as the connecting 
point of the ``wire'' to connect to the opposite electrode through an 
external resistor. We refer to those atoms as ``front atoms.'' They
serve as endpoints for the dedicated split charge that models the external 
resistor. We investigate several different modes of operation of the 
external circuit:

\begin{enumerate}
\item \textbf{switch open, no electrical connection.} The ``external split 
       charge'' is constant. This mode is used for equilibration runs, as 
       well as for aging tests of our battery. 
\item \textbf{switch closed, constant resistance, discharging.} In this case, the current is determined by the 
      instantaneous difference in chemical potential (i.e., the voltage) between the electrodes, divided by a
      fixed resistance. Initially, the voltage is approximately given by the difference of the electrode's electronegativities, 
      modified by the electric field effected by the instantaneous charge distribution.
      If charge transfer continued in this manner indefinitely, the transferred charge would set up an opposing electric 
      field after some time, resisting further charge transfer. However, this electric field causes a charge 
      separation in the electrolyte, and ions rearrange to compensate it. Recall that the separator 
      allows free exchange of electrolyte particles across the half cells. There still is a charge buildup in the
      electrodes, making it energetically favorable for the electrodes to shed some of that charge. This is achieved
      by oxidizing surface atoms on the anode, and releasing them into the solution. 
      Analogously, at the cathode ions dissolved in the electrolyte adsorbing to the surface are reduced.
      This way, the electrodes are neutral again, and the voltage returns (or approaches) that
      of the initial state. This process ends when there are no further ions to be dissolved (and/or adsorbed). 
\item \textbf{switch closed, constant resistance, charging.} This is the same setting as in the discharge case, 
      except that we add an external voltage opposing (and overpowering) the discharge voltage. We investigate to which 
      extent the electrodes return to their previous state, and observe the battery's hysteresis. This allows 
      to study, for instance, surface passivation.
\item \textbf{constant power or constant current.} For brevity, we only present data 
      for discharge under constant resistance, and not under constant power or constant 
      current, even though it is possible to model those discharge modes as well. The 
      \textit{constant power mode} is a good approximation to numerous real-world applications, 
      as many electronic devices need a minimum power throughput to function properly. In the 
      \textit{constant current mode}, charge can continue to flow even beyond the point when the 
      voltage drops to zero. 
      At that point the anode and cathode reverse their roles. This setting allows studying of 
      over-discharging behavior achievable when multiple batteries with differing remaining 
      capacities are connected in series. In that case, the voltage of the cells that still 
      have capacity remaining can drive the empty ones into pole reversal. 
      
\end{enumerate}

\subsection{Unit system and parameters}\label{subsec:units}

\begin{table}
  \caption{Normalized model parameters in our default system.\textsuperscript{\emph{$\dagger$}}}
  \label{table:LJ_parameters}
  \begin{tabular}{lcccc}
\hline\hline
atom type     & $\chi$  & $\kappa$ & description & base charge\\
\hline
$1$            & $-4$  & $4$ & anode atom       & 0\\
$2$            & $4$   & $4$ & cathode atom     & 0\\
$3$            & $-4$  & $4$ & anode cation     & +1\\
$4$            & $4$   & $4$ & cathode cation   & +1\\
$5$            & $1$   & $4$ & electrolyte cation & +1\\
$6$            & $1$   & $4$ & electrolyte anion & -1\\
\hline\hline
 bond type     &  $\varepsilon _{\mathrm{LJ}}$  & $\sigma _{\mathrm{LJ}}$ & $\kappa^{(0)}_{ij}$\textsuperscript{\emph{$\ddagger$}}\\
\hline
$(1,2)-(1,2)$  & $1.75$ & $1.0$ & $3.0$ &\\
$(1,2)-(3,4)$  & $1.0$  & $1.0$ & $3.0$ &\\
$(1,2)-(5,6)$  & $1.0$  & $1.0$ & &\\
$(3,4)-(3,4)$  & $0.75$ & $1.0$ & $3.0$ &\\
$(3,4)-(5,6)$  & $2.0$  & $1.0$ & &\\
$(5)-(5)$      & $0.5$  & $1.0$ & &\\
$(5)-(6)$      & $0.5$  & $1.2$ & &\\
$(6)-(6)$      & $0.5$  & $1.0$ & &\\

    \hline\hline
  \end{tabular}

  \textsuperscript{\emph{$\dagger$}} {see Sec.~\ref{subsec:units} for details on the normalization used.}
  \textsuperscript{\emph{$\ddagger$}} {The parameter $\kappa^{(0)}_{ij}$ is explained in Sec.~\ref{subsec:BondHard}. 
Electrolyte (solvent) atoms do not exchange partial charges in our model, neither among themselves
nor with other species, and therefore those bonds do not have a $\kappa^{(0)}_{ij}$ assigned to them.}
\end{table}

Throughout this work we use dimensionless parameters.
In order to facilitate the interpretation of the presented data, 
this section provides ballpark estimates for representative values 
of real materials. 

The unit of charge can be associated with the elementary charge 
$\left[Q\right] = 1.6\times 10^{-19}~\mathrm{C}$. 
For the unit of length, we choose $\left[l\right] = 2.3~\mathrm{\AA}$. 
This is to approximate the Lennard-Jones parameter $\sigma_{\mathrm{LJ}}$ used for Cu-Cu interactions in 
the literature.\cite{HwangEtAl2004} The value for Zn is comparable in magnitude albeit slightly larger. 
We define the
 unit of mass as the atomic mass of copper, $\left[m\right] \approx 64~\mathrm{amu}$. 
As last independent unit, the energy is normalized to 
$\left[E\right] = 0.16~\mathrm{eV} \approx 15.5~\mathrm{kJ/mol}$. 
Again, this is so that the interaction between metallic electrode particles 
is comparable to values for $\epsilon_{\mathrm{LJ}}$(Cu-Cu) in the literature.

Together, length, charge, energy, and mass specify a complete set of units for our purposes. 
Derived units are the unit of current $\left[I\right] \approx 350~\mathrm{nA}$, 
$\left[R\right] \approx 3.3 \times 10^{5}~\mathrm{\Omega}$ as resistance normalization, 
$\left[U\right] \approx 160~\mathrm{mV}$ as unit of voltage, 
and $\left[t\right] \approx 0.5~\mathrm{ps}$ as unit of time.
Our battery demonstrator operates at a temperature $T \approx 740~\mathrm{K}$, 
comparable to a liquid-salt battery.\cite{BradwellEtAl2012}

With these choices of units, the default value for the electronegativity 
difference between anode and cathode is $\Delta \chi = 1.28$~V, which is close
to many standard cells, e.g., alkaline (1.5~V) or NiMH (1.2~V) batteries.
In the default parameterization, we use atomic hardnesses of
$\kappa = 0.64$~eV, which is much smaller than typical values,
e.g., $\kappa_{\rm Cu} = 7.3$~V/e and $\kappa_{\rm Li} = 4.7$~V/e.\cite{GhoshIslam2010}
Moreover, unlike real systems, our standard values for $\chi$ and $\kappa$ 
do not reproduce a neutral dissociation limit of dimers, 
because $\kappa/2 - \chi > 0$, as can be seen from Eq.~(\ref{eq:SQE_potential}).
We made this choice of parameters to accelerate the generation of ions.
At the same time, we
ensured in selected simulations that the qualitative features of the
discharge curves remained unchanged for much larger values of $\kappa$
and smaller values for $\chi$ (see Sec.~\ref{sec:results}). 

In order to compare our results with macroscopic systems, our model
battery would have to be scaled up by a factor of roughly $10^{21}$. 
In principle, each spatial dimension can have a different scaling factor,
however, for simplicity, one may assume the same factor of
$10^7$ in each direction. 
An inherent ambiguity of how to scale the direction
normal to the interfaces remains.
One way to scale the simulation is to take our nano-battery as an electrical element 
and connect $10^{7}$ of them in series, and $10^{14}$ in parallel. 
This method yields an overall open-circuit voltage (OCV) of $12.8~\mathrm{MV}$
for our default choice of parameters
($\Delta \chi = 8$). 
Connecting the external resistors in a similar fashion as the batteries
leads to a scaled resistance of $66~\mathrm{\Omega}$ 
(microscopic system: $R = 2000$ in dimensionless units).
Consequently, the nominal macroscopic discharge current would be 
$70$~kA.
Discharge now takes $10^7$ times longer (about $0.1$~s) than in 
the microscopic case, because the total number of
transferrable charges increases by $10^{21}$, while the current
only increases by $10^{14}$. 

Alternatively, one can scale both battery and resistor {\it as a whole}
in each spatial dimension. This way one retains the value for the 
macroscopic resistance of $66~\mathrm{\Omega}$. In contrast, the 
voltage in this case still has its microscopic value of 
$1.28~\mathrm{V}$. The resulting nominal macroscopic discharge current 
is $7$~mA, which is not very much lower than real-world currents.
Now the discharge would take $10^{14}$ times longer than in our
microscopic model (about 14~d). We stress that the discharge 
characteristics of such a scaled-up version of our model battery would 
be different from the ones presented in this work because the electrode 
surface-to-volume ratio would be much reduced in the macroscopic battery, 
among other reasons. 

In many figures we show discharge curves, i.e., plots showing the instantaneous
voltage vs. the transferred relative charge. In those plots, the voltage is
normalized by the electrode atom's difference in electronegativities, because that is
the difference in chemical potential in absense of any charge effects or electric 
fields. The transferred relative charge is the total charge flown across the external 
resistor, normalized by the total number of atoms in either of the electrodes. We consider 
not only the dissolvable layers but also the fixed electrode atoms. Each atom can change 
its oxidation state by one: the mobile layers can desorb as cations, while the fixed 
layer's charge is balanced through the formation of a double layer from the 
electrolyte. Note that the charge through the external resistor
is fractional because we implement it as a dedicated (and not-equilibrated) split charge.
We chose to do this instead of sending across only integer increments in order to
get smooth curves. This can be considered in implicit average over many time steps. 

\subsection{Limitations and code efficiency}\label{subsec:limitations}

At this stage, our model only contains some rudimentary
approximation to chemistry, and the ability to model redox reactions. We only consider 
two-body forces, and leave dihedral and torsional interactions for future work. Moreover, at 
short distances we do not screen the Coulomb interaction, even though the wave functions of atoms 
overlap in such a situation, and the point-charge approximation breaks down. 

Our current implementation can be made to run much more efficiently. 
Particular points to note are the following. 
We do not cut off the Lennard-Jones interaction, which makes its complexity $\mathcal{O}(N^{2})$. 
We also compute the Coulomb interactions with an expensive $\mathcal{O}(N^{2})$
direct-sum algorithm, which limits the system sizes we can currently study. 
We could save computing time with a more efficient approach.
We plan an implementation into the open-source code LAMMPS, 
which will alleviate this limitation.

Even with such improvements the solution of the large linear system is more expensive
than the setup of the system matrix, for which the Coulomb term is calculated. Two
ways to make this cheaper come to mind: the first is to use a more efficient algorithm 
for the solution of the linear system, for example using a conjugate gradient 
method.\cite{numericalRecipes} Second, we could update 
only the split charge in vicinity of the ICT. 
At this developmental stage, we re-equilibrate \textit{all} partial charges after an ICT. 
Instead, one could implicate only the partial charges within some cutoff 
radius $R^{\mathrm{opt}}$. This would mean introducing a finite signal speed for the SQE, as a 
split charge could be transferred a certain distance in one time step. The 
instantaneousness of the update would be lost, making it harder to model metals. 
The advantage is that it would make the operation $\mathcal{O}(N)$ instead of 
$\mathcal{O}(N^{2})$. 
The resulting error can be estimated and is bounded: the change in the 
split charges, and thus the error as a result of restricting the update distance,
drops off exponentially with increasing cutoff distance.\cite{NistorMueser2009}

Lastly, in future implementations we will randomize the order of bonds for which we
attempt an ICT and for which an ICTM is carried out, and eliminate the slight
ordering bias currently present in the code.

All of the optimizations described above are left for future work;
even without them (albeit for small systems) the method yields 
encouraging results.

\section{Results}\label{sec:results}

In this section we describe the results gleaned from our simulations. 
We emphasize again that our model is more a proof-of-concept rather than a 
faithful representation of a real battery. As such, the parameters 
have not been chosen appropriate for any specific material. Our intent 
is to demonstrate that redoxSQE nevertheless reproduces generic 
properties of macroscopic batteries without further input.

The model necessarily has a number of parameters, albeit not nearly as many 
as some mesoscopic porous electrode simulations.\cite{MenzelEtAl2011} Some 
are microphysical and chemical parameters, for instance the LJ properties 
of the materials, the atomic hardness, or the electronegativities. Those are 
in principle readily parameterizable or even measurable quantities for real 
materials, but we elect to use representative values (see 
Sec.~\ref{subsec:units}), rather than to model specific materials. 
The parameters associated with the bond hardness are not 
as easily measured, but can in principle be fit to values found from measurements 
and \textit{ab initio} quantum-chemical DFT simulations, such as ESP (electrostatic potential) partial 
charges, Hirshfeld-I charges, and dipole moments of 
molecules.\cite{NistorEtAl2006,Mathieu2007,VerstraelenEtAl2009}
Other parameters are implementation choices such as number of atoms 
to model, the electrode setup, and the geometry of our cell.
Finally, we vary parameters that have a documented and experimentally 
accessible impact on battery performance, such as the temperature, 
and whether the battery is discharged 
continuously or in pulses. The dependence on this last set of parameters 
results naturally and self-consistently with our method, and does not
have to be put in implicitly or explicitly.

\subsection{Dependence on internal model parameters}\label{subsec:results_internal}

In this section, we explore the dependence of the discharge characteristics 
on internal model parameters, while Sec.~\ref{subsec:results_external} 
focuses on external parameters. In SQE, the difference in electronegativity 
between atoms of two metals determines the open-circuit voltage (OCV). In all 
following plots, the voltage is normalized to this value. 

\begin{figure}[hbtp]
  \includegraphics[width=1.0\columnwidth,angle=0]{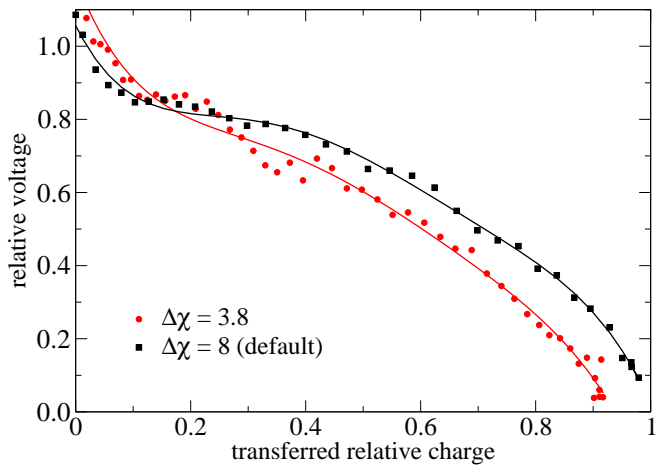}
  \caption{ Color online. Discharge curves of our battery demonstrator for different electronegativities. The 
            data represents an average over several independent runs, and each point is averaged 
            over many MD time steps. The solid lines are inserted to guide the eye. 
            Our default model has $\chi = \pm 4$ and $\kappa=4$ which violates the dissociation limit: dissociating 
            a diatomic molecule will produce ionic instead of neutral products. The runs with 
            $\chi = \pm 1.9$ do not have this flaw, and still exhibit similar discharge behavior, 
            even though the likelihood of anode atoms to dissolve as ions is reduced. The scatter 
            is larger for smaller $\chi$ since the theoretical voltage is reduced by a factor of 
            $2.1$, and thus the signal-to-noise ratio halves, even though in this case we average 
            over 8 different realizations. 
  }
  \label{fig:discharge_comparison_chi}
\end{figure}

Separating a diatomic molecule adiabatically results in neutral products for each pair of stable elements. 
However, if $\Delta \chi > (\kappa_{1}+\kappa_{2})/2$, it is energetically favorable 
that negative integer charge (i.e., an electron) remains on the more electronegative 
partner.\cite{Mueser2012}
For a multi-atom system, the expression for the neutral dissociation limit is not as simple anymore, because the 
atomic hardness is reduced in an ensemble.\cite{Mueser2012} In our default 
system, we choose $\chi=\pm 4$ and $\kappa = 4$, in order to facilitate the 
formation of ions at the anode, even though these values lead to a violation 
of the neutral diatomic dissociation limit. 
However, Fig.~\ref{fig:discharge_comparison_chi} shows that this does not have a 
large impact on the discharge curve: even for $\chi=\pm 1.9$ and $\kappa = 4$, 
satisfying the neutral dissociation limit, the qualitative picture remains. 

We note that the large scatter stems from the small number of
atoms in our simulations. If we dissolve $50\%$ of the total number of anode 
atoms, we have 30 cations in solution, and a statistical uncertainty of 
$\mathcal{O}\left(1/\sqrt{30}\right)$, i.e., $\approx 15\%$. The stochastic 
error reduces by a factor of two by averaging over 4 independent realizations. 

\begin{figure}[hbtp]
  \includegraphics[width=1.0\columnwidth,angle=0]{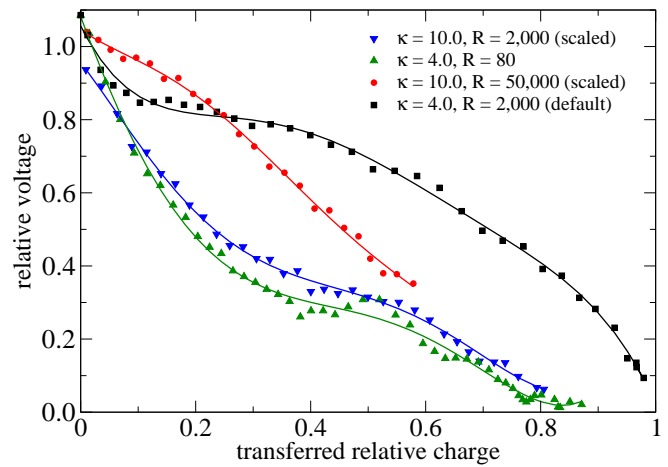}
  \caption{ Color online. Discharge curves of our battery demonstrator varying the atomic hardness. The 
            data represents an average over 4 independent runs, and each point is averaged 
            over many MD time steps. The solid lines are inserted to guide the eye. 
            Runs with $\kappa = 10$ (approximating real atomic hardnesses more closely than 
            our standard value) require much higher external resistances in order to achieve
            similar discharge behavior as for the default value of $\kappa$. For larger values, 
            the plateau is much less pronounced. The $\kappa = 10$ runs are scaled down by a
            factor of 1.3 so that they overlap the $\kappa = 4$ curves better (see text).           
  }
  \label{fig:discharge_comparison_kappaAtomic}
\end{figure}

In Sec.~\ref{subsec:units} we reported representative values as normalization 
for the dimensionless parameters used in this work, and noted that our standard 
value for the atomic hardness $\kappa$ was rather low compared with values for 
real materials. In Fig.~\ref{fig:discharge_comparison_kappaAtomic}, we show the 
discharge behavior of our nano-battery for $\kappa = 10$ and $\Delta\chi = 8$, 
again satisfying the neutral diatomic dissociation limit. In this case, ion 
formation (i.e., redox reactions at the electrode surface) takes much longer, and 
the external resistor needs to be scaled up in order to get similar behavior. A 
factor of 2.5 in $\kappa$ is approximately compensated by a factor of 25 in 
resistance. In addition, the plateau is far less pronounced for the larger 
$\kappa$, and the discharge proceeds faster. Note that the initial voltage in 
this case is also higher. The reason is that some pre-dissolved cations 
adsorb to the electrode immediately, and cause a greater difference in 
chemical potential, and therefore OCV. In order to have the curves overlap for 
better visual comparability, 
we scaled down the results for $\kappa=10$ by a factor of 1.3, which compensates 
for the larger OCV.  Note that the simulation with $R = 50,000$ and $\kappa = 10$ 
takes very long, for the reasons described above, and was terminated before all 
charge was transferred.

\begin{figure}[hbtp]
  \includegraphics[width=1.0\columnwidth,angle=0]{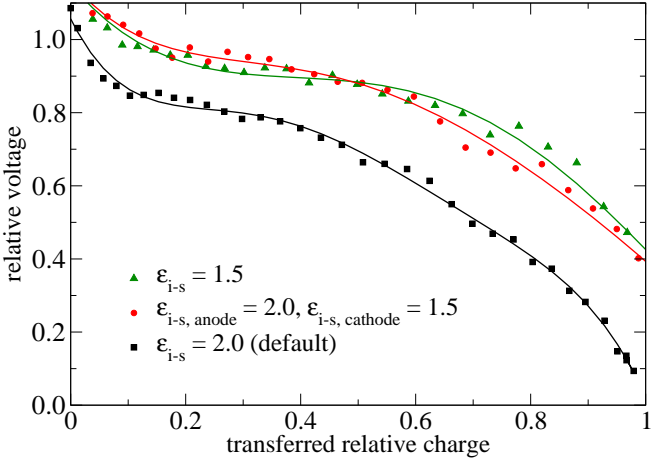}
  \caption{ Color online. Discharge curves of our battery demonstrator varying the LJ parameter 
            $\varepsilon _{\mathrm{LJ}}$ between electrode ions and the solvent (electrolyte). The 
            data represents an average over 4 independent runs, and each point is averaged 
            over many MD time steps. The solid lines are inserted to guide the eye. 
            This figure exemplifies the optimization potential that lies in picking optimal 
            materials for the battery: a smaller affinity between electrolyte and cations boosts 
            battery performance significantly. The anode half-cell reaction seems less important in 
            this respect, as optimizing it does not yield a comparable additional improvement.
  }
  \label{fig:discharge_comparison_epsIS}
\end{figure}

In Fig.~\ref{fig:discharge_comparison_epsIS}, we exemplify the influence of 
the LJ parameters on the discharge curve. We vary 
$\varepsilon _{\mathrm{LJ}}$ between electrode ions and the electrolyte. A 
higher value means that it is favorable for an ion to surround itself with 
electrolyte atoms, as opposed to other atoms with which its 
$\varepsilon _{\mathrm{LJ}}$ is lower. The discharge curve has a higher 
and more extended plateau for a smaller $\varepsilon _{\mathrm{LJ}}$, caused by the reaction 
on the cathode, where a lower value means that it is more likely that an 
ion is adsorbed to the electrode. At the anode, the opposite should be the 
case. There, a \textit{greater} $\varepsilon _{\mathrm{LJ}}$ should make it 
more likely for an ion to be dissolved into the electrolyte. However, the 
test case with 
$\varepsilon _{\mathrm{LJ}} ^{\mathrm{anode}} > \varepsilon _{\mathrm{LJ}} ^{\mathrm{cathode}}$ 
changes the plateau only marginally. We conclude that the cathode reaction 
is more important in this respect. If the method in implemented into more 
sophisticated software (e.g., LAMMPS), it can be used with more realistic force fields for 
particle-particle interactions than what is used herein (i.e., simple two-particle LJ interaction).

\begin{figure}[hbtp]
  \includegraphics[width=1.0\columnwidth,angle=0]{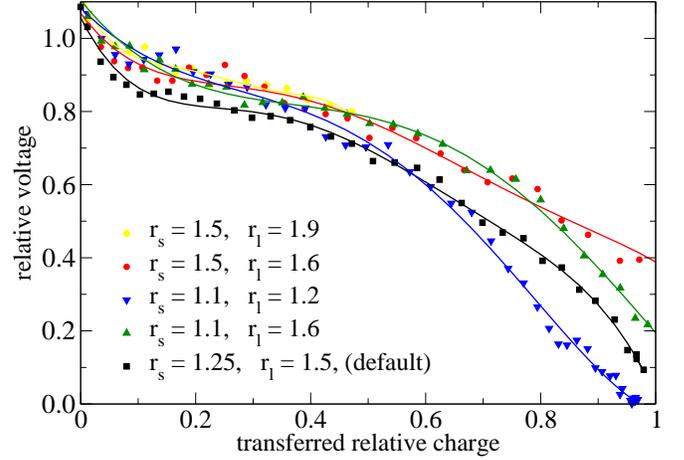}
  \caption{ Color online. Discharge curves of our battery demonstrator for various cutoffs in the parameterizations of 
            the bond hardness. The 
            data represents an average over 4 independent runs, and each point is averaged 
            over many MD time steps. The solid lines are inserted to guide the eye. 
            A change in the cutoffs entering $\kappa _{ij} ^{(\mathrm{b})}$ by $50\%$ does not 
            significantly alter the results: all curves resemble our 
            default model.
            This indicates that the detailed form of $\kappa _{ij} ^{(\mathrm{b})}$ is not of great importance.            
  }
  \label{fig:discharge_comparison_cutoffs}
\end{figure}

\begin{figure}[hbtp]
  \includegraphics[width=1.0\columnwidth,angle=0]{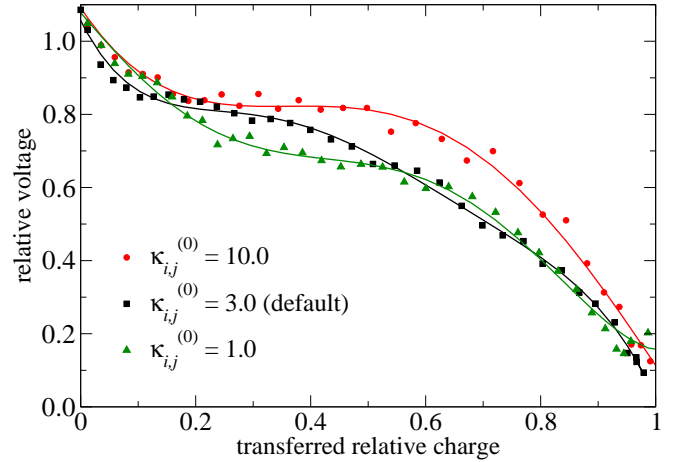}
  \caption{ Color online. Discharge curves of our battery demonstrator for different parameterizations of 
            $\kappa _{ij} ^{(\mathrm{b})}$. The 
            data represents an average over 4 independent runs, and each point is averaged 
            over many MD time steps. The solid lines are inserted to guide the eye. 
            We vary the parameter $\kappa _{ij} ^{(0)}$ by a factor of $3$ to greater and smaller 
            values. The results do not depend sensitively on this choice.
  }
  \label{fig:discharge_comparison_kappa}
\end{figure}

While the LJ parameters in principle can be deduced from experiments, 
pinning down the parameterization details of the bond hardness so that 
they match experiments or \textit{ab initio} results is harder to accomplish. 
\cite{NistorEtAl2006,Mathieu2007,VerstraelenEtAl2009} Fortunately, those 
parameters do not affect the results very strongly, as evidenced in 
Fig.~\ref{fig:discharge_comparison_cutoffs}. A change by $50\%$ in the 
cutoffs for $\kappa _{ij} ^{(\mathrm{b})}$, given in Eq.~(\ref{eq:BondHard}), 
does not make a big difference: all curves nearly overlap with 
those of our default model in the practically relevant regime (until a 
discharge of $\gtrsim 60\%$). In Fig.~\ref{fig:discharge_comparison_kappa}
we scale the parameter $\kappa _{ij} ^{(0)}$ up and down by a factor of $3$. 
The results do not depend sensitively on this choice, 
either.\footnote{Note that the cutoffs and $\kappa _{ij} ^{(0)}$ are not fully
independent. As shown in Eq.~(\ref{eq:BondHard}), all three have a 
scaling effect on $\kappa _{ij} ^{(\mathrm{b})}$.} 
These results indicate that the detailed form of 
$\kappa _{ij} ^{(\mathrm{b})}$ is not of great importance. The only criterion
is that the next-nearest neighbor should not be connected with a dielectric 
bond. This translates into the ``long'' cutoff $r_{\mathrm{l}}$ be 
substantially smaller than the distance to the next-nearest neighbors. 
Otherwise a great number of additional ICTs will be attempted. This does not change 
the result, either, but the computations will be slowed down tremendously.
The simulation shown by the yellow curve in Fig. \ref{fig:discharge_comparison_cutoffs} 
shows this case, and we terminated it before all charge had transferred.

We tested the effect of various other simulation parameters. 
For brevity we only describe the results without including additional figures. 
\begin{itemize}
\item We increased the damping by a factor of 10 without observing any qualitative nor quantitative 
      changes in the discharge behavior. This means that the Langevin damping only has the desired effect to limit
      the battery heating up as a consequence of energy release, but is not strong enough to influence the dynamics much.
\item A variation in the LJ parameters between electrode metals and their ions by $50\%$ also did not alter
      the characteristics significantly. Our choice stems from the idea that ions are bound less tight to the electrodes 
      than neutral metal atoms. This again aids the release of cations into solution.
\item We modified the relative importance of the different contributions to the potential energy. A decrease of the 
      effect of the Coulomb energy by up to $33\%$ (and thereby a corresponding increase in relative importance of
      the other effects) did not influence the discharge curve qualitatively.
\item Similarly, neither scaling the battery in one direction, nor doubling the amount of electrolyte 
      without adding more electrode atoms, nor changing the electrode surface area modifieded the results significantly.
      We conclude that we are not hampered by a lack of electrolyte. However, reducing the number of electrolyte 
      atoms to half its default value will 
      %most likely
      limit the number of ions that can be 
      dissolved, and deteriorate our battery demonstrator's performance.       
\end{itemize}

\subsection{Dependence on external factors, rates and aging}\label{subsec:results_external}

\begin{figure}[hbtp]
  \includegraphics[width=1.0\columnwidth,angle=0]{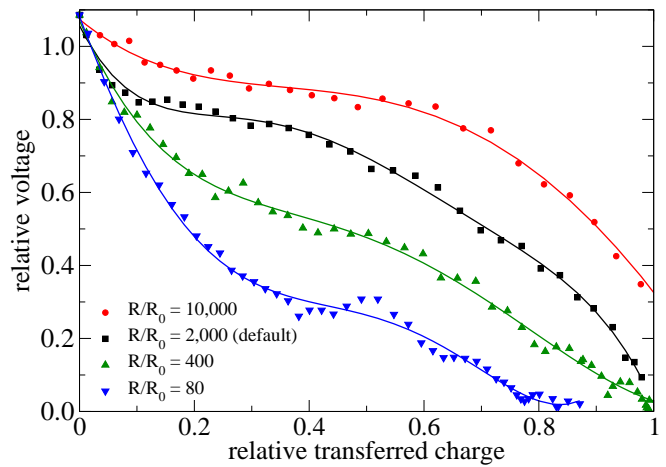}
  \caption{ Color online. Discharge curves of our battery demonstrator with different 
            external resistors. The data represents an average over 4 independent 
            runs, and each point is averaged over many MD time steps. The 
            solid lines are inserted to guide the eye. 
            The higher the external resistance, the closer the battery's behavior 
            approaches the theoretical discharge curve. In this property, and the 
            shape of the discharge curve, our nano-battery resembles a macroscopic 
            battery. Initially, the voltage declines sharply, as the electrodes are 
            charged before ions are dissolved or adsorbed, respectively. This is 
            followed by an extended plateau when the voltage stays constant as the 
            charge transfer through the external resistance is balanced by an equal 
            amount of ion transfer in the electrolyte. At the same time, additional 
            charge on the electrodes is now compensated by dissolving and adsorbing 
            ions. Finally, another steep decline concludes the discharge, as the 
            electrodes are consumed and their surfaces passivated.
  }
  \label{fig:discharge_comparison_resistance}
\end{figure}

An ideal battery should retain its theoretical voltage until the active material has been used up, that is, 
until the anode is completely dissolved, or all free cations have been reduced and adsorbed at the cathode surface. 
Only at this point should the voltage drop to zero.\cite{Linden2010} In reality, batteries have an internal resistance, 
and both the electrolyte and the electrode are polarizable. The former reduces the actual voltage drop across 
the battery, while the latter is responsible for forming Helmholtz double layers at the electrode surfaces, and thus 
depleting some of the battery's capacity.\footnote{A battery's capacity (in Wh) is the integral under the curve
voltage vs. transferred charge, the discharge curve.} Such effects cause both the actual working voltage as well as the usable
capacity to be reduced from their theoretical limits. Additionally, a battery's voltage also depends on the discharge current, 
such that a higher discharge current will decrease the discharge voltage. In Fig.~\ref{fig:discharge_comparison_resistance}
we show curves demonstrating this behavior. At high external resistance (i.e., low discharge current), the voltage stabilizes 
at $\lesssim 90\%$ of its theoretical voltage until about $70\%$ of the available charge has been transferred, at which point
it drops quickly. At medium resistance (our default model), the voltage does not feature such a pronounced plateau, but still 
has $\approx 60\%$ of its voltage at $60\%$ discharge. In contrast, the voltage for a discharge at high currents decreases much
more steeply. These curves are reminiscent of those presented in Refs.~\cite{DoyleEtAl1996,SrinivasanNewman2004,WangSastry2007,Linden2010}, the discharge curves of Panasonic's 
zinc carbon batteries,\cite{Panasonic2009} and those of Duracell's alkaline batteries.\cite{Duracell2012}
If the internal resistance of a battery exceeds the external resistance, we effectively have a short circuit, and the battery 
will discharge as a capacitor, with an initial exponential decay of the voltage.

\begin{figure}[hbtp]
  \includegraphics[width=1.0\columnwidth,angle=0]{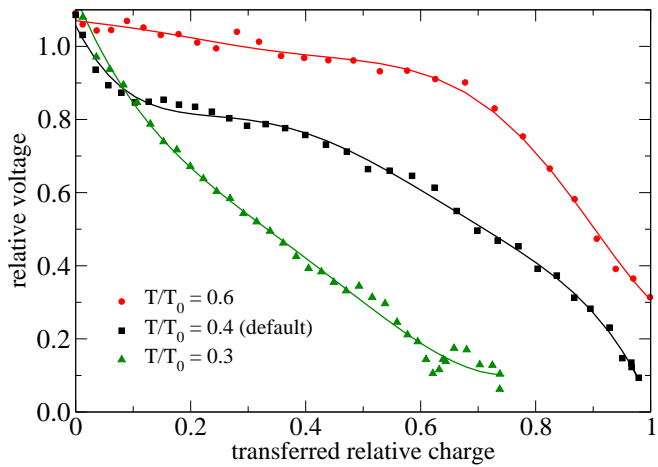}
  \caption{ Color online. Discharge curves for different system temperatures. The 
            data represents an average over 4 independent runs, and each point is averaged 
            over many MD time steps. The solid lines are inserted to guide the eye. 
            At lower temperatures ($T \approx 550~\mathrm{K}$), the battery's performance deteriorates 
            dramatically, while it improves at a higher temperature ($T \approx 1,100~\mathrm{K}$).
  }
  \label{fig:discharge_comparison_temp}
\end{figure}

A factor that is of crucial importance to
real batteries is the temperature at which they operate. Electric vehicles need to be able to reliably operate at
temperatures ranging from $\lesssim -30^{\circ}~\mathrm{C}$ all the way to $\gtrsim 50^{\circ}~\mathrm{C}$. However, low temperatures
decrease both the actual voltage as well as the battery's capacity.\cite{Linden2010}
Figure~\ref{fig:discharge_comparison_temp} shows the temperature dependence of our battery demonstrator. As 
in macroscopic batteries, the voltage is closer to the theoretical voltage for high temperatures, while it is significantly
reduced for lower temperatures. The battery's capacity --- the area under the curve --- decreases by $\approx 42\%$ if the 
temperature is lowered by $25\%$, and increases by $\approx 46\%$ if the temperature is raised by $50\%$ from our default 
temperature. 
We note that our temperature of $T=740~\mathrm{K}$ is quite large, comparable to liquid-salt 
batteries,\cite{BradwellEtAl2012} but greatly exceeds room-temperature. However, we once more 
emphasize the qualitative nature of our findings, not their realism. 

\begin{figure}[hbtp]
  \includegraphics[width=1.0\columnwidth,angle=0]{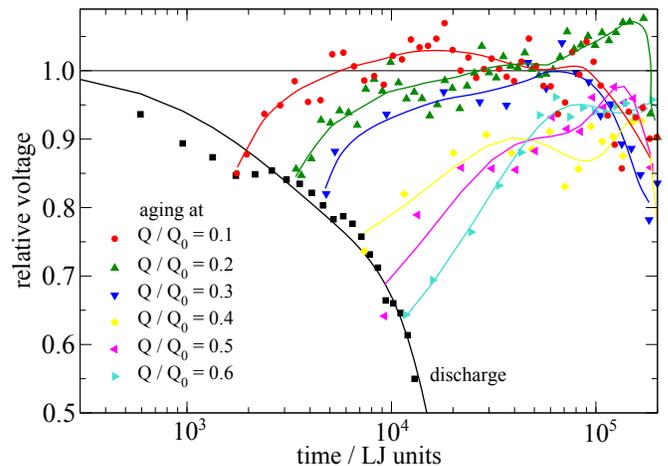}
  \caption{ Color online. Aging studies of our model battery. The 
            data represents an average over 4 independent runs, and each point is averaged 
            over many MD time steps. The solid lines are inserted to guide the eye. 
            After a certain amount of charge has flown through the resistor, the switch is opened
            and the relaxation observed. In all cases, the voltage recuperates to almost the nominal 
            value (with fluctuations). This behavior would not be present in a capacitor, which does
            not recover its charge after the circuit is broken. After a certain time, ions pass the 
            separator and cause self-discharge, which gives rise to a voltage drop. 
  }
  \label{fig:aging}
\end{figure}

In Fig.~\ref{fig:aging} we present aging studies of our model battery. It is common in some electronic 
devices to intersperse recuperation periods with discharge periods. During this time, polarization 
effects are reduced and some of the initial voltage can be recovered.\cite{Linden2010} In order to 
examine the effects of intermittent discharge on our model battery, we discharge it until a certain
amount of current has been drained, and then open the switch and let the system age. Were it an 
electrochemical capacitor, no recovery of voltage would be expected. However, we see a nearly full
recuperation of the initial voltage (with some fluctuations). This voltage is held for $\approx 10^{7}$
time steps. After some time, ions manage to pass through the separator (which is a thermally 
activated process, see Sec.~\ref{subsec:Separator}). This causes a voltage drop that contributes 
to self-discharge. In a macroscopic battery this process will not take place as quickly as in our 
nanoscale device, but one mechanism for is elucidated in our model. The use of redoxSQE furthermore 
allows us to study the morphology changes that the electrodes have undergone, how much surface 
material is passivated and other microphysical parameters of interest.

\begin{figure}[hbtp]
  \includegraphics[width=1.0\columnwidth,angle=0]{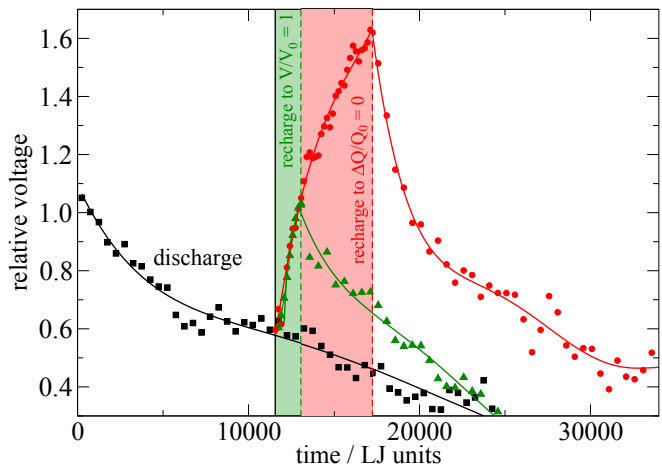}
  \caption{ Color online. Discharge and recharge curves of our battery demonstrator. 
            The lines are inserted to guide the eye.
            Our default system is discharged (black squares) at constant resistance, 
            until $\approx 60\%$ of the total charge has been transferred through the 
            external resistor. At this point, 
            a constant external charging current is switched on; its magnitude is chosen 
            approximately three times the (average) discharge current. This charging 
            current is switched off again when the voltage reaches the OCV 
            (middle, green triangles), or when all charge has been transferred back (top, 
            red circles). In the latter case, the voltage is much higher than the initial OCV.
  }
  \label{fig:charging}
\end{figure}

All reactions occurring in our system are micro-reversible, therefore our nano-battery is a 
secondary, rechargeable cell. Figure~\ref{fig:charging} shows what happens if we recharge 
our battery demonstrator. We discharge our default system at the default constant resistance 
($R=2000$). After $\approx 60\%$ of the total capacity (35 out of 59 integer charges) have 
been transferred through the external circuit, a charging current is switched on, in opposite 
direction as the discharge current. We choose its magnitude approximately three times the 
``average'' discharge current (averaged over the complete discharge of the same model). 

We consider two cases: in the first, the charging current is 
switched off again when the voltage reaches $\approx 1.1$ times the OCV. In this case, the discharge curve in 
the second cycle has deteriorated compared with the initial discharge; the battery has degraded.
The reason is that the electrodes do not fully return to their initial state during the 
charging, but merely the electrolyte reconfigures to balance the dissolved cations in either half cell.
If one were to let the system relax after charging, some dissolved cations would return to the anode,
and some material deposited on the cathode would also dissolve again. A realistic all-atom simulation
using redoxSQE can be useful in the investigation of dendrite formation in this process which can 
short-out and destroy Li-ion batteries.\cite{TarasconArmand2001,ArmandTarascon2008,BhattacharyyaEtAl2011}
It could also help to study the cycling behavior of batteries, and their degradation.

In the second case, all charge is transferred back. Then, the voltage is much higher than the initial 
OCV. This behavior is also seen in real batteries,\cite{Linden2010} and stems
from a buildup of a polarization layer opposite that which forms during discharge. In this case, the 
second discharge is not very dissimilar from the first one.

\section{Discussion and conclusions}\label{sec:discussion}

In this work, we demonstrated that the redoxSQE method~\cite{Mueser2012} can 
model redox reactions in an atomistic molecular dynamics setting, and used it 
to simulate a nanoscale battery demonstrator. 
Even though we did not use a parameterization describing any real energy materials, 
we reproduced generic discharge curves of macroscopic batteries. For example, lower operating 
temperatures reduce the effective capacity of a battery. Higher discharge rates have the 
same effect, but the voltage recuperates when the battery is aged (e.g., discharged in 
pulses). Upon recharging, the battery performance degrades slightly, and the electrode
surface morphology changes during the battery's operation.

Some internal model parameters are not fully accessible experimentally, 
such as the bond hardness, which is an ad-hoc parameter arrived at in a top-down fashion,
by describing the bond-breaking behavior. In a theoretical work, Verstraelen 
\textit{et al.}~\cite{VerstraelenEtAl2013} connect the bond hardness to parameters 
computed with atom-condensed DFT, i.e., derive it in a bottom-up fashion, and may help
to motivate this parameter and ascertain its value quantitatively. 
We showed that, for the battery demonstrator, the results neither depend strongly on 
the detailed implementation, nor on the precise value of the bond hardness.

The \textit{atomic} hardness plays an important role in determining the time scales 
of ion formation at the electrode interface, and thus determines partly the internal 
resistance of the battery, while the electronegativity difference sets the open-circuit 
voltage. But changing those two parameters does not alter the qualitative discharge picture. 
Rather, the battery's behavior is predominantly defined by external 
quantities such as temperature, rates of discharge. We obtain 
results similar to the intermittent discharge mode,\cite{Linden2010} and see self-discharge
when the battery is aged too long. 

In a recent previous paper we applied the same technique to case studies in 
contact electrification between two clusters of ideal metals and ideal 
dielectrics, respectively, showcasing its ability to simulate 
history-dependence.\cite{DappMueser2013a} RedoxSQE reproduces charge 
hysteresis effects during approach and retraction.

One shortcoming of our current implementation is that the electrolyte 
is modeled with fixed-charge particles that do not participate in split charge exchange, nor in ICTs. 
It is an ideal insulator for electrons, and the lack of electronic conduction leaves 
only penetration of the separator by ions as self-discharge mechanism. This idealization
will be abandoned in future work.  

Further development effort will need to be expended on optimizing the method 
(see Sec.~\ref{subsec:Setup}, and Ref.~\cite{Mueser2012}) to make multi-million 
atom simulations possible. An implementation into LAMMPS is planned. 
In order to simulate specific materials or battery setups (such as alkaline batteries,
or Li-ion rechargeables), much chemically-specific parameterization will need to
be done.\cite{NistorEtAl2006,Mathieu2007,VerstraelenEtAl2009,VerstraelenEtAl2012}
Furthermore, more realistic empirical many-body force fields are necessary
for realistic all-atom simulations. 
We point out that the model in its current implementation can best describe
non-directed interactions, as they are prevalent for instance in Alkali batteries, 
with their isotropic reactions of s-orbitals.

Notwithstanding those necessary improvements, it is encouraging that the method
already reproduces generic features of batteries. Mesoscale battery models require 
many assumptions and intimate knowledge of the materials in question, and cannot 
answer fundamental microphysical questions. DFT/MD methods, on the other hand, 
have been used for highly detailed and isolated problem aspects, but need to stay 
away from the electrode-electrolyte interface where redox reactions take place. 
Arguably, this is the most interesting region, as it determines not only 
the ultimate cell performance, but also is where cell degradation takes place.
Harris \textit{et al.}~\cite{HarrisEtAl2010} write ``\textit{the ability to predict 
cell degradation remains a challenge because so many unaccounted for and 
seemingly unrelated micro-scale degradation mechanisms have been identified 
or postulated. [...] Without ... theoretical analysis, 
cause-and-effect relationships between observation and degradation 
pathway can be difficult to demonstrate.}''
RedoxSQE is a first step toward filling this gap. It allows to model all aspects
of a (microscopic) battery in one simulation, and gather insights into the processes
happening at the electrode-electrolyte interface. 

Besides modeling an entire all-atom battery, redoxSQE could also serve as 
part of hybrid multiscale schemes, where 
bulk phenomena inside an electrode or within the electrolyte are computed with a 
mesoscopic model while the electrochemical activity is tackled by redoxSQE.

Daniel's Handbook of Battery Materials~\cite{DanielBesenhard2011} states as 
requirement for a generic life estimation model that it ``\textit{must relate the measured 
cell performance at any given time to a combination of ... effects.}'' In conjunction with 
a more realistic force field and with a proper parameterization of the 
materials, redoxSQE holds promise to enable the study of battery degradation 
and the optimization of battery performance.

\begin{acknowledgments}
We thank R. Nistor
and Y. Qi for useful discussions,
and the J\"ulich Supercomputing Centre for computing time.
\end{acknowledgments}

\bibliographystyle{aipnum4-1}
% \bibliography{bibliography.bib}

\begin{thebibliography}{65}
\bibitem [{Noteb()}]{Noteb}%
  \BibitemOpen
  \bibinfo {note} {Note that the cutoffs and $\kappa _{ij} ^{(0)}$ are not
  fully independent. As shown in Eq.~(\ref {eq:BondHard}), all three have a
  scaling effect on $\kappa _{ij} ^{(\protect \mathrm {b})}$.}\BibitemShut
  {Stop}%
\bibitem [{Notec()}]{Notec}%
  \BibitemOpen
  \bibinfo {note} {A battery's capacity (in Wh) is the integral under the curve
  voltage vs. transferred charge, the discharge curve.}\BibitemShut {Stop}%
\end{thebibliography}


\end{document}